\documentclass[twoside,twocolumn,9pt]{article}
\usepackage{extsizes}
\usepackage{sectsty}
\usepackage[super,sort&compress,comma]{natbib}
\usepackage[version=3]{mhchem}
\usepackage[left=1.5cm, right=1.5cm, top=1.785cm, bottom=2.0cm]{geometry}
\usepackage{balance}
\usepackage{textcomp}
\usepackage{gensymb}
\usepackage{mathptmx}
\usepackage{lmodern}
\usepackage{graphicx}
\usepackage{lastpage}
\usepackage[format=plain,justification=justified,singlelinecheck=false,font={stretch=1.125,small,sf},labelfont=bf,labelsep=space]{caption}
\usepackage{float}
\usepackage{fancyhdr}
\usepackage{fnpos}
\usepackage[english]{babel}
\addto{\captionsenglish}{%
  
}
\usepackage{array}
\usepackage{droidsans}
\usepackage{charter}
\usepackage[T1]{fontenc}
\usepackage[usenames,dvipsnames]{xcolor}
\usepackage{setspace}
\usepackage[compact]{titlesec}
\usepackage{stackengine}
\usepackage{braket}
\newcommand{\op}[1]{\mathop{}\!\ensurestackMath{\stackon[-.95ex]{%
  {#1}}{\smash{{\hat{}}}}}}
\newcommand{\opdag}[1]{\mathop{}\!\op{#1}^{\dag}}

\usepackage{epstopdf}

\definecolor{cream}{RGB}{222,217,201}

\begin{document}

\pagestyle{fancy}
\thispagestyle{plain}
\fancypagestyle{plain}{
\renewcommand{\headrulewidth}{0pt}
}

\makeFNbottom
\makeatletter
\renewcommand\LARGE{\@setfontsize\LARGE{15pt}{17}}
\renewcommand\Large{\@setfontsize\Large{12pt}{14}}
\renewcommand\large{\@setfontsize\large{10pt}{12}}
\renewcommand\footnotesize{\@setfontsize\footnotesize{7pt}{10}}
\makeatother

\renewcommand{\thefootnote}{\fnsymbol{footnote}}
\renewcommand\footnoterule{\vspace*{1pt}%
\color{cream}\hrule width 3.5in height 0.4pt \color{black}\vspace*{5pt}}
\setcounter{secnumdepth}{5}

\makeatletter
\renewcommand\@biblabel[1]{#1}
\renewcommand\@makefntext[1]%
{\noindent\makebox[0pt][r]{\@thefnmark\,}#1}
\makeatother
\renewcommand{\figurename}{\small{Fig.}~}
\sectionfont{\sffamily\Large}
\subsectionfont{\normalsize}
\subsubsectionfont{\bf}
\setstretch{1.125} 
\setlength{\skip\footins}{0.8cm}
\setlength{\footnotesep}{0.25cm}
\setlength{\jot}{10pt}
\titlespacing*{\section}{0pt}{4pt}{4pt}
\titlespacing*{\subsection}{0pt}{15pt}{1pt}

\fancyfoot{}
\fancyfoot[LO,RE]{\vspace{-7.1pt}\includegraphics[height=9pt]{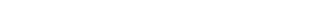}}
\fancyfoot[CO]{\vspace{-7.1pt}\hspace{13.2cm}\includegraphics{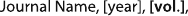}}
\fancyfoot[CE]{\vspace{-7.2pt}\hspace{-14.2cm}\includegraphics{RF}}
\fancyfoot[RO]{\footnotesize{\sffamily{1--\pageref{LastPage} ~\textbar  \hspace{2pt}\thepage}}}
\fancyfoot[LE]{\footnotesize{\sffamily{\thepage~\textbar\hspace{3.45cm} 1--\pageref{LastPage}}}}
\fancyhead{}
\renewcommand{\headrulewidth}{0pt}
\renewcommand{\footrulewidth}{0pt}
\setlength{\arrayrulewidth}{1pt}
\setlength{\columnsep}{6.5mm}
\setlength\bibsep{1pt}

\makeatletter
\newlength{\figrulesep}
\setlength{\figrulesep}{0.5\textfloatsep}

\newcommand{\topfigrule}{\vspace*{-1pt}%
\noindent{\color{cream}\rule[-\figrulesep]{\columnwidth}{1.5pt}} }

\newcommand{\botfigrule}{\vspace*{-2pt}%
\noindent{\color{cream}\rule[\figrulesep]{\columnwidth}{1.5pt}} }

\newcommand{\dblfigrule}{\vspace*{-1pt}%
\noindent{\color{cream}\rule[-\figrulesep]{\textwidth}{1.5pt}} }

\makeatother

\twocolumn[
  \begin{@twocolumnfalse}
{\includegraphics[height=30pt]{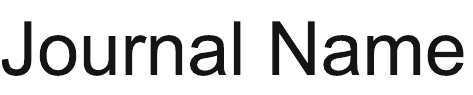}\hfill\raisebox{0pt}[0pt][0pt]{\includegraphics[height=55pt]{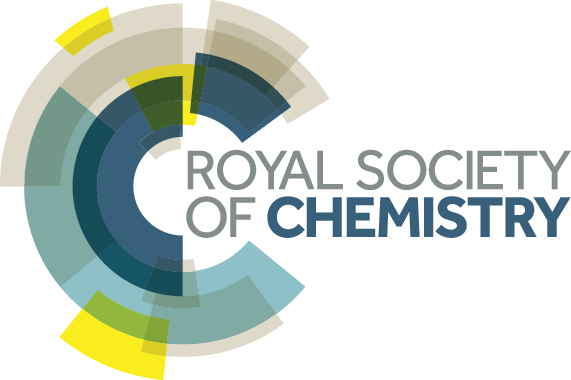}}\\[1ex]
\includegraphics[width=18.5cm]{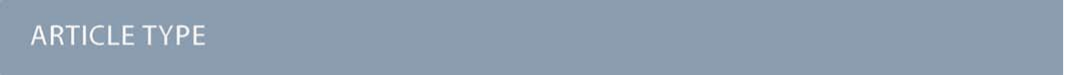}}\par
\vspace{1em}
\sffamily
\begin{tabular}{m{4.5cm} p{13.5cm} }

\includegraphics{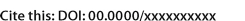} & \noindent\LARGE{\textbf{Molecular Origins of Exciton Condensation in Van der Waals Heterostructure Bilayers$^\dag$}} \\
\vspace{0.3cm} & \vspace{0.3cm} \\

 & \noindent\large{Lillian I. Payne Torres ,\textit{$^{a}$} Anna O. Schouten,\textit{$^{a}$} and David A. Mazziotti\textit{$^{a}$}} \\

\includegraphics{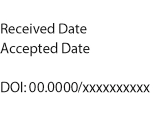} & \noindent\normalsize{Recent experiments have realized exciton condensation in bilayer materials such as graphene double layers and the van der Waals heterostructure MoSe$_2$-WSe$_2$ with the potential for nearly frictionless energy transport. Here
we computationally observe the microscopic beginnings of exciton condensation in a molecular-scale fragment of MoSe$_2$-WSe$_2$, using advanced electronic structure methods based on reduced density matrices.  We establish a connection between the signature of exciton condensation---the presence of a large eigenvalue in the particle-hole reduced density matrix---and experimental evidence of exciton condensation in the material. The presence of a ``critical seed'' of exciton condensation in a molecular-scale fragment of a heterostructure bilayer provides insight into how local short-range strongly correlated effects may give rise to macroscopic exciton condensation. We find that molecular-scale properties such as layer alignment and interlayer distance can impact the formation of nonclassical long-range order in heterostructure bilayers, demonstrating the importance of geometric considerations for the rational design of exciton condensate material.  Mechanistic insights into the microscopic origins of exciton condensation have potential implications for the design and development of new materials with enhanced energy transport properties.} \\
\end{tabular}

 \end{@twocolumnfalse} \vspace{0.6cm}

  ]

\renewcommand*\rmdefault{bch}\normalfont\upshape
\rmfamily
\section*{}
\vspace{-1cm}


\footnotetext{\textit{$^{a}$~Department of Chemistry and The James Franck Institute, The University of Chicago, Chicago, IL 60637}}

\footnotetext{\dag~Supplementary Information available: XYZ coordinates of the 7 bilayers. See DOI: 00.0000/00000000.}


\section*{Introduction}

Current power transmission technologies are plagued by energy losses due to frictional dissipation of energy, and the search for materials capable of frictionless energy transport at ambient pressure and temperature is ongoing. Excitons, quasi-bosonic bound states of electrons and holes, are capable of quantum condensation. The resulting superfluid is theoretically capable of nondissipative energy transfer,\cite{fil_electron-hole_2018,keldysh_coherent_2017} which could inspire novel electronic devices and spur tremendous innovation to efficient energy transfer applications. Additionally, exciton condensation is predicted to be possible at higher temperatures than traditional superconductivity.\cite{fuhrer_chasing_2016} While the condensation is difficult to achieve because excitons are susceptible to recombination, especially at room temperature, it has been experimentally realized through the coupling of excitons to polaritons,\cite{kasprzak_boseeinstein_2006,fuhrer_chasing_2016} and the spatial separation of electrons and holes in bilayer materials.\cite{min_room-temperature_2008, kogar_signatures_2017,li_excitonic_2017,wang_evidence_2019} Bilayer systems provide an important platform for exciton condensation due the spatial separation of electrons and holes between layers preventing quick recombination of excitons. Graphene bilayers have proven to be promising candidates for exciton condensation, with the twist-angle dependence of their electronic states affording an avenue for tuneability.\cite{min_room-temperature_2008,li_excitonic_2017,hu_quantum-metric-enabled_2022,liu_quantum_2017,su_spatially_2017} Recently, there has been experimental evidence for exciton condensation in bilayers composed of transition metal dichalcogenides such as MoSe$_2$-WSe$_2$ heterostructures.\cite{wang_evidence_2019,sigl_signatures_2020,ma_strongly_2021,fogler_high-temperature_2014} Additionally, recent theoretical work has shown the possibility for the beginnings of exciton condensation in several small-scale molecular systems.\cite{safaei_quantum_2018,schouten_exciton_2021,sager_beginnings_2022,schouten_potential_2023,sager_preparation_2020}

 Here we computationally observe exciton condensation emerging in molecular-scale fragments of a family of transition metal dichalcogenide bilayers. Using variational 2-RDM theory,\cite{mazziotti_reduced-density-matrix_2007,mazziotti_contracted_1998,nakata_variational_2001,mazziotti_realization_2004,gidofalvi_active-space_2008,mazziotti_two-electron_2012,mazziotti_pure-_2016,mazziotti_large-scale_2011,cioslowski_many-electron_2000,zhao_reduced_2004,cances_electronic_2006,shenvi_active-space_2010,mazziotti_structure_2012,verstichel_variational_2012,piris_global_2017} we calculate the populations of the excitonic states---the eigenvalues of the particle-hole reduced density matrix---for molecular-scale fragments of several transition metal dichalcogenide bilayers. In the noninteracting limit, all eigenvalues of the modified particle-hole RDM are equal to zero or one, and no condensation is present. In the strongly correlated limit, off-diagonal long-range order emerges in the particle-hole reduced density matrix as a consequence of long-range coherence, leading to eigenvalues greater than one. The presence of a maximal exciton population $\lambda_G$ greater than one, indicating more than one particle-hole pair in a given excitonic mode, is thus the theoretical signature of exciton condensation.\cite{safaei_quantum_2018, garrod_particle-hole_1969}
Exciton condensation requires both stable, long-lived excitons (such as interlayer excitons) as well as strong correlation for long range order in the material. Factors that disrupt the formation of stable excitons or the coherence of the system will disrupt the exciton condensate and cause a decrease in $\lambda_G$, making $\lambda_G$ ideal for probing the effects of geometric modifications to an exciton condensate.

 We find that this signature begins to appear in fragments of MoSe$_2$-WSe$_2$, a material which has previously shown experimental evidence for exciton condensation.\cite{wang_evidence_2019, sigl_signatures_2020, ma_strongly_2021} The degree of condensation increases with the size of the molecular fragment, suggesting that we are seeing a ``critical seed'' of the condensation that would be present in a bulk system. Additionally, we show that layer alignment and interlayer distance influence the extent of exciton condensation. Finally, we find that the signature is also present in several transition metal dichalcogenide bilayers similar to MoSe$_2$-WSe$_2$. The presence of a ``critical seed'' suggests that local, short-range strongly correlated effects are significant for the formation of a macroscopic condensate. Furthermore, our results highlight the importance of geometric considerations in the design of exciton condensate materials. The presence of the signature of exciton condensation in several different van der Waals heterostructure bilayers indicates that the identity of the particular transition metal dichalcogenide does not have a large effect on the potential for exciton condensation, suggesting that materials capable of exciton condensation may be more ubiquitous than previously thought.
\section*{Results}
\begin{figure}
    \centering
    \includegraphics[width=8cm]{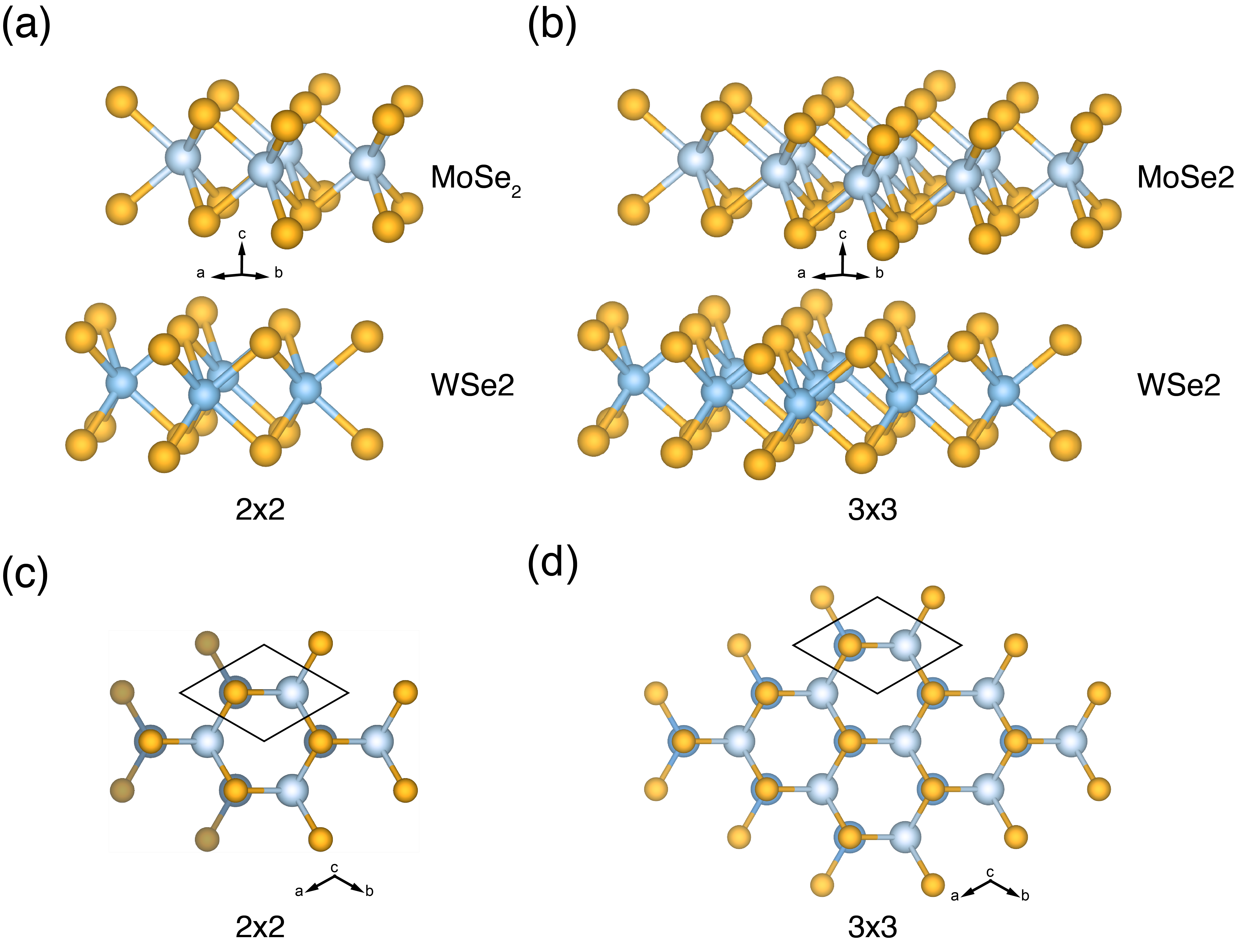}
    \caption{Example $2\times2$ and $3\times3$ cluster models for the MoSe$_2$-WSe$_2$ heterobilayer, with molybdenum atoms in light blue, tungsten atoms in darker blue, and selenium atoms in orange. Passivating hydrogens omitted for clarity. Black outline indicates one primitive cell.}
    \label{fig:models}
\end{figure}

The transition metal dichalcogenide bilayer MoSe$_2$-WSe$_2$ is probed for the signature of exciton condensation---$\lambda_G > 1$, the number of excitons condensed into a single particle-hole state. Two cluster models are constructed for each system, consisting of bilayers with the primitive cell tiled $2\times2$ and $3\times3$ in the XY plane respectively (Fig.~\ref{fig:models}). The signature of exciton condensation is obtained for each molecular structure via variational two-electron reduced density matrix theory (v2RDM) as described in the Methods.\cite{mazziotti_reduced-density-matrix_2007,mazziotti_contracted_1998,nakata_variational_2001,mazziotti_realization_2004,gidofalvi_active-space_2008,mazziotti_two-electron_2012,mazziotti_pure-_2016,mazziotti_large-scale_2011,cioslowski_many-electron_2000,zhao_reduced_2004,cances_electronic_2006,shenvi_active-space_2010,mazziotti_structure_2012,verstichel_variational_2012,piris_global_2017} The theoretical density of the excitonic mode corresponding to the largest eigenvalue is plotted to determine if the system is exhibiting the desired interlayer exciton. To gauge the influence of layer alignment and interlayer distance on the extent of exciton condensation, $\lambda_G$ is calculated for a variety of different layer alignments and interlayer spacings. Finally, a family of related transition metal dichalcogenide bilayers is probed for the signature of exciton condensation.
\subsection*{Exciton condensation in  MoSe$_2$-WSe$_2$}
\begin{figure}
    \centering
    \includegraphics[width=6cm]{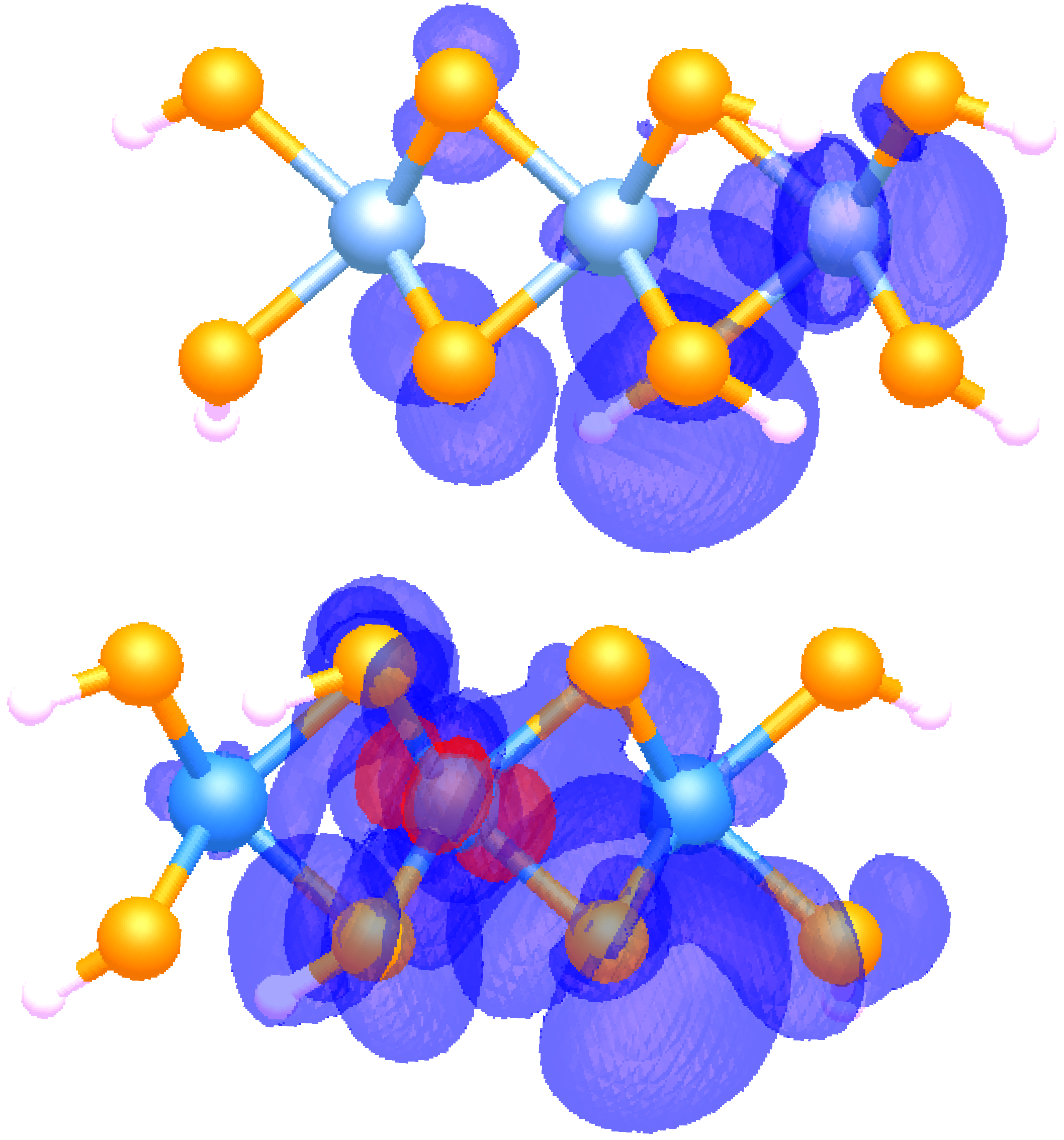}
    \caption{Visualization of exciton density, where the red atomic orbital indicates the constrained location of the hole and the purple indicates the probabilistic density of the electron corresponding to that particular hole location.}
    \label{fig:Mo-WSe2_explot}
\end{figure}

MoSe$_2$-WSe$_2$ is a transition metal dichalcogenide bilayer that has shown strong experimental evidence of exciton condensation.\cite{wang_evidence_2019,sigl_signatures_2020,ma_strongly_2021} Calculation of the largest eigenvalue of the particle-hole reduced density matrix for MoSe$_2$-WSe$_2$ for the $2\times2$ and $3\times3$ models results in a $\lambda_G$ of 1.29 and 1.40, respectively. The presence of $\lambda_G>1$ indicates the potential for exciton condensation, aligning with previous experimental studies. While the exciton population in a single quantum state is quite small, the increase in $\lambda_G$ as system size increases from the $2\times2$ to $3\times3$ model suggests that exciton population will continue to increase as the system approaches the size of a bulk material sample, potentially leading to macroscopic exciton condensation.
To determine if the exciton condensate character is corresponding to the desired interlayer exciton, we visualize exciton density by calculating the probabilistic location of an electron corresponding to a specified hole location (see Methods for more information). The exciton density corresponding to $\lambda_G$ is plotted in Fig. \ref{fig:Mo-WSe2_explot}. Hole density is constrained to a tungsten 5$d_{x^2-y^2}$ orbital (in red) while the probabilistic location of the electron is plotted in purple.

Plotting the exciton density shows both inter- and intra-layer contributions to the exciton mode corresponding to the large eigenvalue. As the interlayer exciton is desirable due to being less prone to quick recombination, the presence of an interlayer exciton state supports the potential for practical realization of exciton condensation in this particular transition metal dichalcogenide bilayer. These results align with results from periodic electronic structure calculations with the Bethe-Salpeter equation, a common method for studying excitons in materials. Calculations of the excitonic spectra with the Bethe-Salpeter equation reveal strong interlayer excitons, with the lowest energy excitonic state composed of holes located on the WSe$_2$ layer and electrons located on the MoSe$_2$ layer.\cite{torun_interlayer_2018,gao_interlayer_2017,sadecka_inter-_2022,gillen_interlayer_2018,li_strain_2023} Excitons with intralayer character are also present. Plots of the excitonic wavefunctions reveal highly delocalized excitons,\cite{torun_interlayer_2018,gillen_interlayer_2018} similar to our observation of delocalized exciton density.

As the molecular models are passivated with hydrogens, it is important to determine if the added hydrogens have a significant impact on the excitonic states we are investigating. The probabilistic density corresponding to the excitonic ground state (and thus the condensate) shows essentially no density residing on the hydrogens, demonstrating that the hydrogen atoms contribute very minimally, if at all, to the exciton state containing the condensate character and thus their inclusion does not have significant impact on the results presented here.
\subsection*{Interlayer distance in MoSe$_2$-WSe$_2$}
Experimental studies of exciton condensation in transition metal dichalcogenide bilayers primarily utilize devices constructed with two transition metal dichalcogenide monolayers separated by one or more layers of hexagonal boron nitride (hBN).\cite{wang_evidence_2019,ma_strongly_2021,kalt_excitonic_2024} The hBN layers are included as a tunnel barrier to prevent exciton recombination. Previous theoretical work has suggested that interlayer distance can have a large effect on the extent of exciton condensation in bilayer systems,\cite{sager_beginnings_2022,schouten_potential_2023} suggesting that varying the thickness of the tunnel barrier could serve as a way to tune exciton condensation in transition metal dichalcogenide bilayers.

To explore the influence of varying interlayer distance between the transition metal dichalcogenide layers on the extent of exciton condensation, $\lambda_G$ is calculated for MoSe$_2$-WSe$_2$ ($2\times2$) with interlayer distances ranging from 2.745 to 5.745 \AA. This distance range was chosen to be in the realm of those achievable with hBN layers, although computational expense prohibits the inclusion of actual hBN layers.

\begin{figure}
    \centering
    \includegraphics[width=8cm]{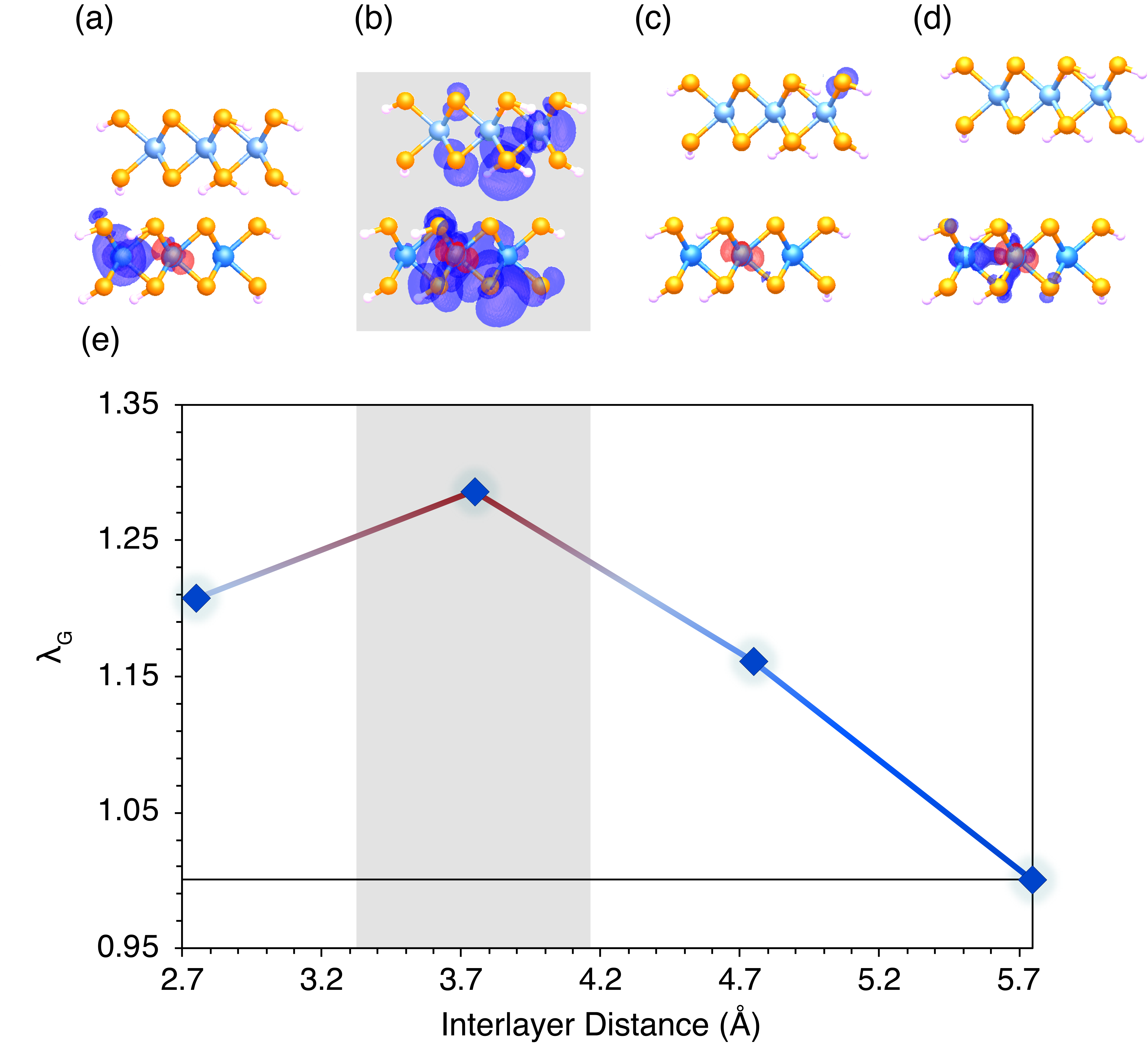}
    \caption{Visualization of the exciton density of the exciton mode corresponding to the largest eigenvalue of the particle-hole density matrix ($\lambda_G$) (a-d) for different interlayer spacings of MoSe$_2$-WSe$_2$. The largest eigenvalue corresponding to each density is plotted below each image against the distance between MoSe$_2$ and WSe$_2$) layers (e). The atomic orbital corresponding to the constrained location of the hole is plotted in red, while the probabilistic density of the electron is plotted in purple. The density plot and data point for the original interlayer distance are highlighted in grey.}
    \label{fig:spacing}
\end{figure}

These results, plotted in Fig. \ref{fig:spacing}e, show the signature of exciton condensation maximized around  3.745 {\AA} (the layer spacing of the MoSe$_2$-WSe$_2$ model used in the calculations above). Directly above the data point for each interlayer distance is plotted exciton density of the exciton mode corresponding to that particular large eigenvalue (Fig. \ref{fig:spacing}a-d). These plotted exciton densities show interlayer exciton character and electron delocalization maximized around 3.745 {\AA}, the interlayer distance which also maximized $\lambda_G$. As interlayer distance increases to 4.745 {\AA}, an interlayer exciton is barely present, and the electron density is highly localized on a selenium p orbital. At 5.745 {\AA}, all interlayer character has vanished and the exciton is localized on adjacent tungsten d orbitals. When interlayer distance is decreased to 2.745 {\AA}, all interlayer exciton character has once again vanished and the exciton is localized on adjacent tungsten d orbitals as for the 5.745 {\AA} case. These calculations appear to demonstrate that interlayer distance impacts the formation of the interlayer exciton desirable for long exciton lifetimes, which is a necessary factor for the formation of an exciton condensate. It should be noted that these calculations do not include the actual hBN spacer that would be used in experimental devices, and so the exact distance trends observed here may not be consistent to those seen experimentally. We believe, however, that the general trend observed in our calculations will hold in the experimental case---that there is a range of ideal layer separation, and distances above and below that range will result in a decrease in observed exciton condensation.
\subsection*{Layer alignment in MoSe$_2$-WSe$_2$}
\begin{figure}
    \centering
    \includegraphics[width=8cm]{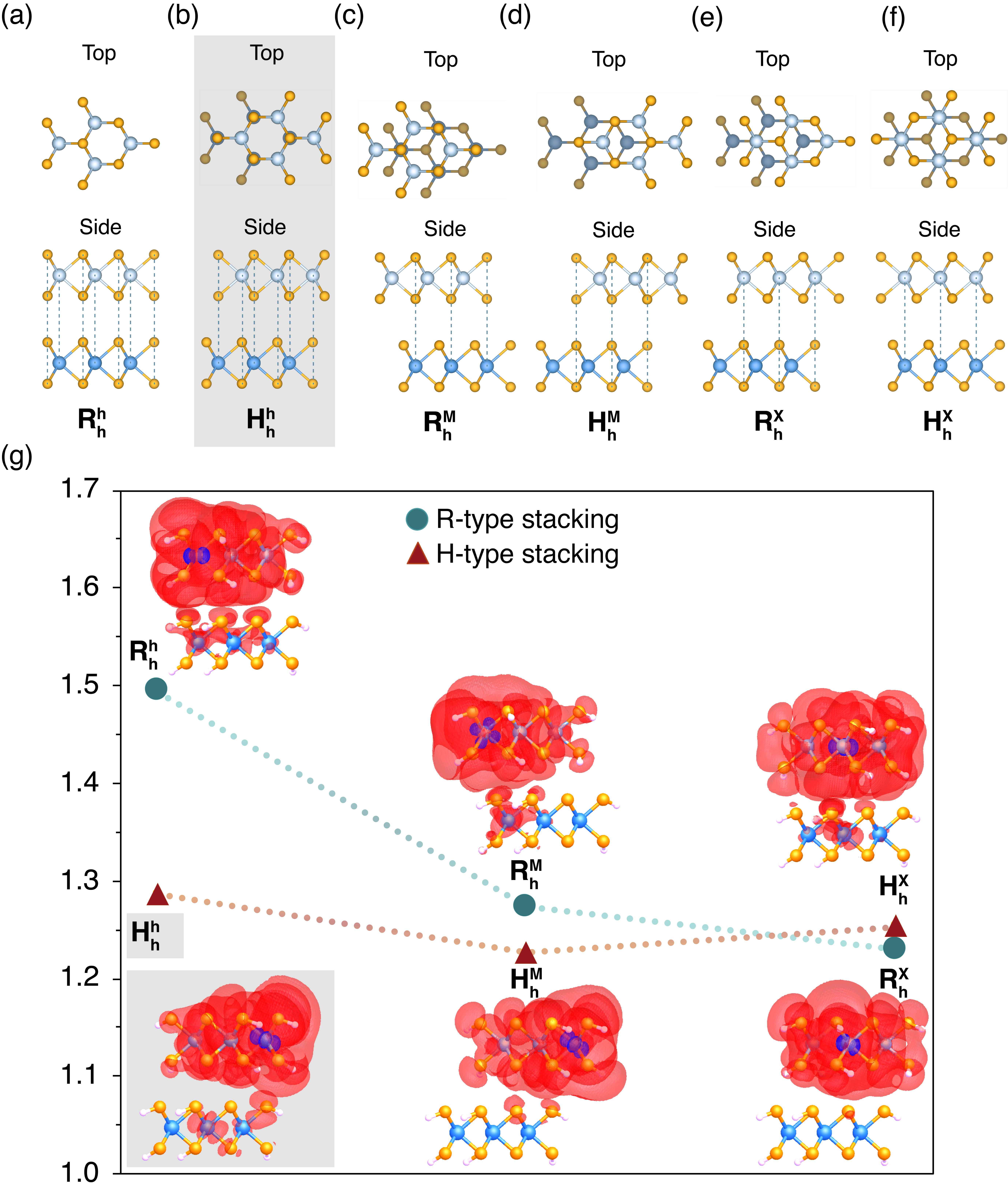}
    \caption{Schematics of six high-symmetry stacking orders of MoSe$_2$-WSe$_2$ (a-e). Passivating hydrogens omitted for clarity. The largest eigenvalue of the particle-hole reduced density matrix ($\lambda_G$) is plotted for all high symmetry stackings, with R-type stackings plotted with blue circles and H-type stakings plotted with red triangles for clarity. Visualizations of the exciton density of the exciton mode corresponding to $\lambda_G$ are provided next to each data point. The atomic orbital corresponding to the constrained location of the particle is plotted in purple, while the probabilistic density of the hole is plotted in red. The schematic and data point corresponding to the original layer alignment ($H^h_h$) are highlighted in grey.}
    \label{fig:rotate}
\end{figure}

The 2D nature of the heterostructure bilayers means that electronic properties can be engineered via different interlayer stacking alignments.\cite{he_stacking_2014,jiang_valley_2014,wang_stackingengineered_2021,hu_stacking_2016,wang_mirror_2019}
Due to the broken inversion symmetry in TMD bilayers, two distinct stacking orders emerge: a parallel configuration termed R (rhombohedral, $0\degree$ alignment between the layers) and an antiparallel configuration termed H (hexagonal, $180\degree$ alignment between the layers). Each configuration gives three high-symmetry stacking orders.
To assess the influence of interlayer alignment on the degree of exciton condensation, $\lambda_G$ is calculated for $2\times2$ models based on all six high-symmetry stacking orders of MoSe$_2$-WSe$_2$. These six stacking orders are referred to via the terminology used in Ref.\cite{zhao_excitons_2023}, wherein the superscript and subscript indicate to the top and bottom layers respectively, and \textit{h}, \textit{X}, and \textit{M} refer to the hexagonal center, the transition metal atom, and the chalcogen atom. Each model is constructed by rotating the MoSe$_2$ layer of MoSe$_2$-WSe$_2$ such that the layer alignment corresponds to one of the six high symmetry stacking orders of MoSe$_2$-WSe$_2$. The signature of exciton condensation is plotted for each of the six interlayer alignments in Fig. \ref{fig:rotate}g, with the exciton density of the mode corresponding to the large eigenvalue plotted next to each point. Note that here the location of the electron (purple) is constrained while the probabilistic density of the hole (red) is plotted. The location of the constrained particle differs between stackings, as the atomic-orbital composition of the active molecular orbitals involved in the exciton differs between stackings. A schematic of each alignment is depicted in Fig. \ref{fig:rotate}a-f.

Calculating exciton population for each layer alignment shows that exciton condensate character is maximized for the $R^h_h$ stacking, and minimized for the $R^X_h$ and $H^M_h$ stackings (although the eigenvalues of all stackings other than $R^h_h$ are quite similar, falling in the range of 1.23-1.29. These differences in $\lambda_G$ are likely due to differing interlayer interactions between the different stacking orders. Differing layer alignments lead to differing electrostatic interactions between atomic sites,\cite{he_stacking_2014,marom_stacking_2010} which may impact exciton formation. Differing orbital overlap may also drive the differences in $\lambda_G$ observed--aligning with previous experimental and theoretical studies which have highlighted the impact of interlayer twist angle, and thus interlayer coupling and hybridization, on exciton formation in TMD van der Waals heterostructures.\cite{shimazaki_strongly_2020,liu_electronic_2015,forg_moire_2021,jiang_interlayer_2021,zhang_twist-angle_2020} It is notable that the $R^h_h$ stacking, which corresponds to each metal atom aligned directly over a metal atom on the opposite layer and each chalcogen atom directly over a chalcogen atom on the opposite layer, results in greater exciton condensation relative to the other stackings. This suggests that maximizing the alignment between sites on opposing layers may be significant for the emergence of large $\lambda_G$.

By comparing the exciton density of the mode corresponding to the large eigenvalue of the different high symmetry stackings, it is apparent that while each stacking has highly delocalized \textit{intralayer} exciton density, only the $R^h_h$, $H^h_h$, $R^M_h$, and $H^X_h$ stackings have significant \textit{interlayer} exciton densities. Amongst these, the high-$\lambda_G$ $R^h_h$ stacking has the most interlayer character, with the density of the hole delocalized among all four Tungsten atoms. The $H^M_h$ and $R^M_h$ stackings, which have no major interlayer exciton density, also have the lowest values of $\lambda_G$. These results suggest that the differing interactions between layers in the various stackings make some stackings more favorable for the formation of stable, highly delocalized interlayer excitons, which in turn supports the emergence of exciton condensation.

While it is not clear if this exact trend for the six high symmetry stackings persists in bulk samples of this material, we expect that layer alignment will have a significant impact on exciton formation and thus the ability of the material to form an exciton condensate.

\subsection*{Exciton condensation in different bilayers}
\begin{figure}
    \centering
    \includegraphics[width=8cm]{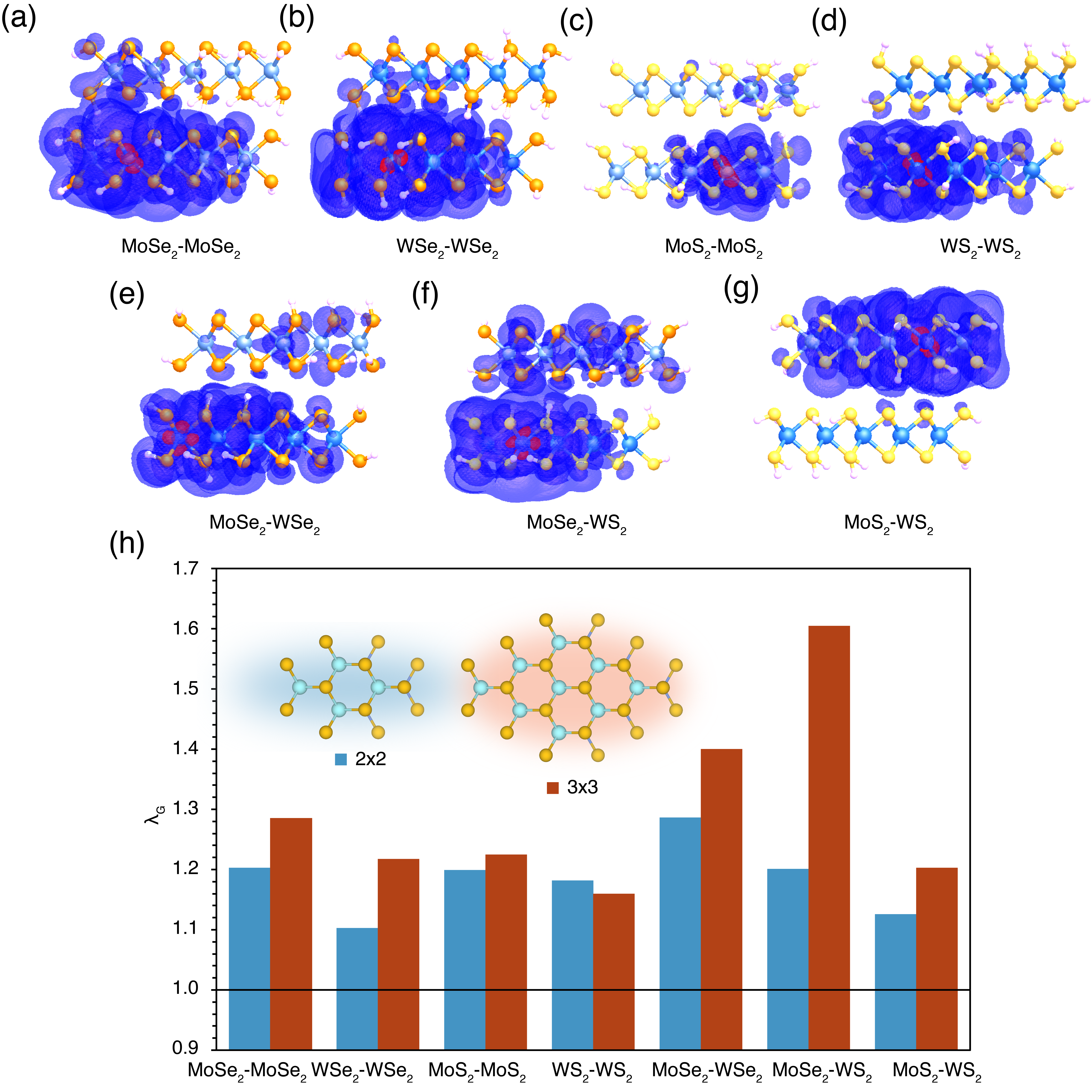}
    \caption{(a-g) Visualizations of the exciton density of the exciton mode corresponding to $\lambda_G$ for the $3\times3$ model of each heterostructure. The atomic orbital corresponding to the constrained location of the hole is plotted in red, while the probabilistic density of the electron is plotted in purple. The largest eigenvalue of the particle-hole density matrix ($\lambda_G$) for both the $2\times2$ and $3\times3$ cluster models of several related homo and heterobilayers. Schematics of a top view of the $2\times2$ and $3\times3$ models are provided for reference (passivating hydrogens omitted for clarity).}
    \label{fig:2x2_3x3}
\end{figure}

It is of interest to see if other TMD bilayers have show potential for the exciton condensation observed in MoSe$_2$-WSe$_2$. To this end, a family of related homo and heterobilayers composed of different combinations of MoSe$_2$, WSe$_2$, Mo$S_2$, and WS$_2$ were chosen for investigation. To probe the chosen transition metal dichalcogenide bilayers for the beginnings of exciton condensation, $\lambda_G$ is calculated for both the $2\times2$ and $3\times3$ models of each bilayer. These results are plotted in Fig. \ref{fig:2x2_3x3}h. The exciton density of the exciton mode corresponding to $\lambda_G$ is plotted for each $3\times3$ model.

As can be seen in Fig. \ref{fig:2x2_3x3}, all homobilayer and heterobilayer systems exhibit $\lambda_G>1$ and thus demonstrate the potential for exciton condensation. The relative magnitudes of $\lambda_G$ between the systems are not necessarily indicative of their relative capabilities of exciton condensation, as the above results have shown that structural factors such as layer spacing have considerable impact on the beginnings of exciton condensation. While all models have interlayer distances between 3.0 and 4.0 {\AA}, close to the ideal layer separation for MoSe$_2$-WSe$_2$ determined previously, the ideal layer separation for these new systems is likely different than that for MoSe$_2$-WSe$_2$.  However, the plotted exciton densities in Fig. \ref{fig:2x2_3x3}a-g give some insight into the trends observed in Fig. \ref{fig:2x2_3x3}h. MoSe$_2$-WS$_2$ displays a notably higher $\lambda_G$ than the other bilayers, which, as in the other cases discussed in the work, is accompanied by an increase in interlayer exciton density. All of the bilayers show highly delocalized exciton density with some interlayer component, which is a necessary factor for the emergence of exciton condensation. The fact that $\lambda_G > 1$ is observed for each new bilayer system, even in very small fragments of the bulk systems speaks promisingly to these material's potential for exciton condensation.

\section*{Conclusions}

Here we observe the microscopic origins of exciton condensation in a family of transition metal dichalcogenide bilayers. We find that the signature of exciton condensation is present in a fragment of MoSe$_2$-WSe$_2$, aligning with the experimental evidence for exciton condensation in this material. Visualizing the exciton density of the exciton mode corresponding to the largest eigenvalue shows that the probabilistic electron density (with respect to a hole constrained to a tungsten 5$d_{x^2-y^2}$ orbital) is highly delocalized, with both intra and interlayer components. The presence of interlayer exciton density aligns with the experimental results, which specifically target the formation of an interlayer exciton condensate. The degree of condensation increases for a larger-sized molecular fragment, indicating the presence of a ``critical seed'' of exciton condensation that may become a macroscopic condensate in a bulk system. We also find that layer alignment and interlayer distance influence the extent of exciton condensation. Out of six high-symmetry alignments of MoSe$_2$-WSe$_2$, the $R^h_h$ alignment shows the largest signature of exciton condensation, likely due to favorable electrostatic interactions and orbital overlap that aid the formation of robust interlayer excitons. Varying interlayer distance between 2.745 and 5.745 {\AA} shows that maximal exciton condensation occurs at an intermediate distance of 3.745 {\AA}, while bringing the layers too close together or too far apart diminishes the extent of exciton condensation. Finally, we find that the signature of exciton condensation is also present in several transition metal dichalcogenide bilayers similar to MoSe$_2$-WSe$_2$, suggesting the potential for exciton condensation. While the magnitudes of $\lambda_G$ differ between the systems, this is not necessarily indicative of their
relative capabilities of exciton condensation, as ideal layer alignment and interlayer separation was not taken into consideration. However, as calculated the MoSe$_2$-WS$_2$ structure produces a robust interlayer exciton with a notably large $\lambda_G$.

While it has been understood that the large eigenvalue signature is a necessary condition for long-range order and thus the beginnings of exciton condensation,\cite{garrod_particle-hole_1969,safaei_quantum_2018} previous work has primarily focused on novel molecular systems that are not direct analogues to those explored experimentally.\cite{schouten_potential_2023, sager_beginnings_2022, schouten_exciton_2021} In this work, by theoretically probing a molecular fragment of a material, MoSe$_2$-WSe$_2$, that has already shown experimental evidence for exciton condensation, we are able to draw a direct connection between the necessary condition for the beginning of exciton condensation and actual experimental observation of condensation. While the large eigenvalues we see in our molecular-scale fragments do not necessarily suggest condensation in the macroscopic sense, we believe they show us the \textit{beginnings} of exciton condensation which may grow into macroscopic exciton condensation as the system grows to approach the thermodynamic limit. This idea is supported by the fact that $\lambda_G$ increases as we increase system size from the $2\times2$ to the $3\times3$ model. Our results also give us unique insight into the molecular origins of the phenomena that is observed in material-scale samples, as the presence of a ``critical seed'' of condensation in a very small fragment of the material, which can be influenced by geometric modifications of the molecular models, suggests that local short-range strongly correlated effects may be significant for the formation of a macroscopic condensate. The fact that layer alignment and interlayer distance impact the magnitude of the signature of exciton condensation also highlights the importance of geometric considerations in the design of exciton condensate materials.  We also observe the signature of exciton condensation in several other transition metal dichalcogenide bilayers in addition to MoSe$_2$-WSe$_2$, suggesting that the selection of transition metal dichalcogenide involved may not have a large impact on the potential for exciton condensation.  This observation may imply that materials capable of exciton condensation are more ubiquitous than previously thought, and motivates further investigation into new bilayer systems for realizing exciton condensation.

\section*{Methods}

\subsection*{Theory}
A definitive theoretical signature for exciton condensation has recently been developed from reduced density matrix (RDM) theory: a large eigenvalue in the particle-hole reduced density matrix

\begin{equation}
^2G^{i,j}_{k,l} = \bra{\Psi}   \opdag{a}_i\op{a}_j\opdag{a}_l\op{a}_k \ket{\Psi}
\label{eq:1}
\end{equation}

\noindent where $\ket{\Psi}$ is the N-electron wavefunction and $\opdag{a}_i$ and $\op{a}_i$ are the fermionic creation and annihilation operators.\cite{safaei_quantum_2018,garrod_particle-hole_1969,kohn_two_1970} The particle-hole RDM can be obtained from the two-electron reduced density matrix (2-RDM) by a linear mapping given by:

\begin{equation}
^2G^{i,j}_{k,l}= \delta^{j}_{l} {}^{1}D^{i}_{k} - {}^{2}D^{i,l}_{k,j}
\label{eq:2}
\end{equation}

\noindent where ${}^{2}D^{i,j}_{k,l}$ is the 2-RDM, $\delta^{i}_{j}$ is the delta function, and $^{1}D^{i}_{k}$ is the one-electron reduced density matrix (1-RDM). Additionally, one must subtract from the particle-hole RDM the components corresponding to a ground-state-to-ground-state transition, which otherwise creates one extraneous large eigenvalue unrelated to exciton condensation:

\begin{equation}
^2\Tilde{G}^{i,j}_{k,l}={}^2G^{i,j}_{k,l}-{}^{1}D^{i}_{j}{}^{1}D^{l}_{k}
\label{eq:3}
\end{equation}

\noindent The largest eigenvalue of this modified particle-hole RDM is the signature of exciton condensation, and is referred to as $\lambda_G$. The eigenvalues of the modified particle-hole reduced density matrix correspond to the occupation of each particle-hole state, and so the magnitude of $\lambda_G$ describes the extent of condensation.
\subsection*{Visualization of Exciton Density}
Visualization of the excitonic state corresponding to $\lambda_G$ is obtained by constraining the location of the excitonic hole to a particular atomic orbital and plotting the probabilistic density of the electron. A matrix of molecular orbitals in terms of atomic orbitals ($M_{MO,AO}$) is obtained from the output of v2RDM calculation performed in the Maple Quantum Chemistry Package. This matrix is then used to calculate a matrix of molecular orbitals in terms of atomic orbitals, $M_{AO,MO}$
\begin{equation}
M_{AO,MO}=(M^{T}_{MO,AO})^{-1}.
\end{equation}
\noindent A submatrix $M^{active}_{AO,MO}$ corresponding to the active orbitals is isolated from $M_{AO,MO}$, and the eigenvector or the modified particle-hole RDM corresponding to $\lambda_G$ is reshaped as a matrix in the basis of the active molecular orbitals. This matrix, called $V_{max}$, is multiplied along with the submatrix
\begin{equation}
(M^{active}_{AO,MO})(V_{max})(M^{active}_{AO,MO})^T
\end{equation}
\noindent to create a matrix representing hole atomic orbitals in terms of the coefficients of the contributions of the electrons to other molecular orbitals.

\subsection*{Molecular Models}
Several different transition metal dichalcogenide bilayers are chosen for investigation. These include homobilayers composed of two layers of MoSe$_2$, WSe$_2$, MoS$_2$, and WS$_2$ respectively, as well as the heterobilayers MoSe$_2$-WSe$_2$, MoS$_2$-WS$_2$, and MoSe$_2$-WS$_2$. In order to make models that can be explored via current electronic structure techniques for strong correlation at reasonable computational cost, cluster models are built from bulk experimental crystal structures retrieved from the Inorganic Crystal Structure Database,\cite{bergerhoff_inorganic_1983,champion_properties_1965,kalikhman_crystal_1972,schonfeld_anisotropic_1983,van_arkel_uber_1926} with dangling bonds saturated by hydrogens. Two cluster models are constructed for each system, consisting of bilayers with the primitive cell tiled $2\times2$ and $3\times3$ in the XY plane respectively. As the primitive cells of the models do not contain a C3 rotational axis, the molecular models do not exhibit the C3 rotational symmetry that may be found in the bulk material. The layers of all heterobilayer structures were spaced arbitrarily at 3.474 {\AA} apart. Molecular coordinates for all structures are available in the Supporting Information.

The 2-RDM is obtained directly from the molecular structure via
variational two-electron reduced density matrix theory (v2RDM).\cite{mazziotti_reduced-density-matrix_2007,mazziotti_contracted_1998,nakata_variational_2001,mazziotti_realization_2004,gidofalvi_active-space_2008,mazziotti_two-electron_2012,mazziotti_pure-_2016,mazziotti_large-scale_2011,cioslowski_many-electron_2000,zhao_reduced_2004,cances_electronic_2006,shenvi_active-space_2010,mazziotti_structure_2012,verstichel_variational_2012,piris_global_2017} We specifically utilize v2RDM-CASSCF, a combination of v2RDM with the complete active space self-consistent-field (CASSCF) method which allows for optimization of orbitals in the chosen active space.\cite{gidofalvi_active-space_2008}. An [8,8] active space is used for all $2\times2$ structures, while an [18,18] active space is used for all $3\times3$ structures, where [X,Y] indicates an active space of X electrons in Y orbitals. These active spaces correspond to an active electron in one orbital on each of the metal atoms for both structures. The def2-SVP basis set is used for all calculations.\cite{weigend_balanced_2005,eichkorn_auxiliary_1997} The particle-hole RDM is then constructed from the 2-RDM via the method described above. The eigenvalues of the particle-hole RDM are then calculated, with the largest of which referred to as $\lambda_G$.

\section*{Author contributions}
D.M. conceived of the project.  L.P. performed theoretical and computational calculations.  L.P., A.S., and D.M. interpreted the results and data and wrote the manuscript.

\section*{Conflicts of interest}
There are no conflicts to declare.

\section*{Data availability}
Data will be provided by the corresponding author upon reasonable request.

\section*{Acknowledgements}

D.A.M. gratefully acknowledges support from the U. S. National Science Foundation Grant No. CHE-2155082, the Department of Energy, Office of Basic Energy Sciences, Grant DE-SC0019215, and the ACS Petroleum Research Fund Grant No. PRF No. 61644-ND6.


\balance


\bibliography{VdWs_references} 
\bibliographystyle{rsc} 
\end{document}